# *Characterization of band offsets in $Al_xIn_{1-x}As_ySb_{1-y}$ alloys with varying Al composition*

Jiyuan Zheng,[1] Andrew H. Jones,[1] Yaohua Tan,[2] Ann K. Rockwell,[3] Stephen March,[3] Sheikh Z. Ahmed,[4] Catherine A. Dukes,[5] Avik W. Ghosh,[4] Seth R. Bank,[3] and Joe C. Campbell,[1,a)]

[1]*Electrical and Computer Engineering Department, University of Virginia, Charlottesville, Virginia 22904, USA*

[2]*Synopsys Inc, 455 N Mary Ave, Sunnyvale, CA 94085, USA*

[3]*Microelectronics Research Center, University of Texas, Austin, Texas 78758, USA*

[4]*Department of Physics, University of Virginia, Charlottesville, Virginia 22904, USA*

[5]*Materials Science and Engineering, University of Virginia, Charlottesville, Virginia 22904, USA*

The unprecedented wide bandgap tunability (~1 eV) of $Al_xIn_{1-x}As_ySb_{1-y}$ latticed-matched to GaSb enables the fabrication of photodetectors over a wide range from near-infrared to mid-infrared. In this paper, the valence band-offsets in $Al_xIn_{1-x}As_ySb_{1-y}$ with different Al compositions are analyzed by tight binding calculations and X-ray photoelectron spectroscopy (XPS) measurements. The observed weak variation in valence band offsets is consistent with the lack of any minigaps in the valence band, compared to the conduction band.

$Al_xIn_{1-x}As_ySb_{1-y}$ materials latticed-matched to GaSb substrate can be utilized to fabricate photodetectors that operate from the near-infrared to mid-infrared photodetectors owing to a wide compositional tuning of the bandgap energy. Previously a miscibility gap[1-3] prevented the development of $Al_xIn_{1-x}As_ySb_{1-y}$ devices with high Al concentrations. Vaughn, *et al.* demonstrated a technique to circumvent this limitation with digital alloy structures with Al fraction up to 35% by using periodic cells comprised of a few monolayers of the binary materials, AlAs, AlSb, InAs, and InSb.[4,5] Maddox, *et al.,* extended this method to cover the entire direct bandgap range of

---

a) Electronic address: jcc7s@ virginia.edu

compositions (to $x \approx 80\%$).[3] This digital alloy material has been used to fabricate a staircase avalanche photodiode (APD)[6] and separate absorption, charge, and multiplication (SACM) APDs that operate at 1550 nm with excess noise comparable to that of Si.[7] To date, there have been few studies of the material characteristics of $Al_xIn_{1-x}As_ySb_{1-y}$. For electronic and optoelectronic devices, the relative conduction and valence band offsets at the interfaces of different compositions are important parameters. In fact, a possible origin of low excess noise in some digital alloys seems to be the emergence of sizeable minigaps inside one carrier band, which may eventually be connected to the variation in corresponding band offsets.[8] Recent electroreflectance measurements on $Al_xIn_{1-x}As_ySb_{1-y}$ indicate that the valence band offsets are very low.[9] In this paper, a first-principles study based on an empirical tight binding model and X-ray photoelectron spectroscopy (XPS) are used to determine the bandgap discontinuities.

Figure 1 shows the periodic structures of the digital alloy $Al_xIn_{1-x}As_ySb_{1-y}$ considered in this paper. Figures 1(a) and 1(b) show the structures for the compositions $x=30\%$ and $x=70\%$, respectively. The $Al_xIn_{1-x}As_ySb_{1-y}$ digital alloy is fabricated by periodically stacking 4 binary materials: InAs, InSb, AlSb, and AlAs; each period consists of 10 mono-layers (ML). The Al composition, $x$, and As composition, $y$, are controlled by the thicknesses of the binary layers. In this paper, the band structures of $Al_xIn_{1-x}As_ySb_{1-y}$ with $x$ varying from 0 to 80% are analyzed. The layer structure of a unit cell for each composition is provided in Table I. The thicknesses of the layers are designed to achieve a lattice match to GaSb substrate.

An environment-dependent empirical tight-binding model is used to calculate the band structure. In this model, neighboring atoms, bond angles, and bond lengths are considered; the empirical parameters were obtained by iteratively adjusting with hybrid density functional (HSE06) results. This model has been verified in many different material structures including the InAlAs digital alloy,[2,8] group IV and III-V heterojunctions, ultrathin Si and MoS$_2$ layers.[10-12] The empirical parameters have been published in the literature.[11] Since the lattice has periodicity along the growth direction, supercells are chosen consistent with the smallest repeatable units. Since we need to maintain the total thickness of each period at 10 ML, the monolayer numbers for the individual layers are not necessarily integers, which means that the adjacent layers overlap. This raises the issue that the lateral distribution of atoms is somewhat random. In order to reduce the calculation time, we simplify the lateral atomic distribution to be regular and thus reduce the size of the supercell.

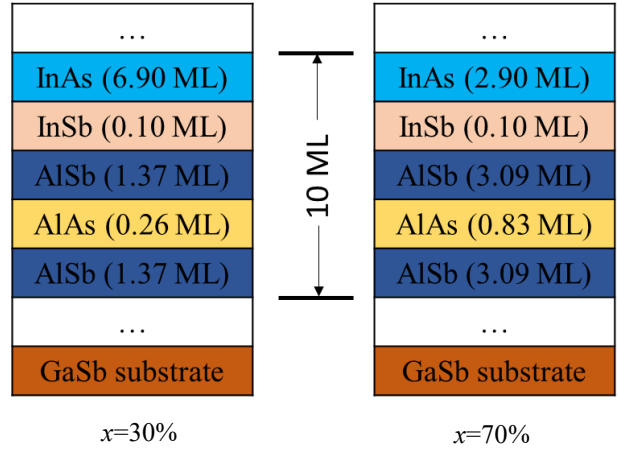

FIG. 1. Lattice structure for (a) $x$=30% and (b) $x$=70% Al$_x$In$_{1-x}$As$_y$Sb$_{1-y}$ digital alloy.

TABLE I. Al$_x$In$_{1-x}$As$_y$Sb$_{1-y}$ digital alloy monolayer fractions for various binary alloys constitutents in a unit cell, for x varying from 0 to 80%.

|  | $x$=0% | $x$=10% | $x$=20% | $x$=30% | $x$=40% | $x$=50% | $x$=60% | $x$=70% | $x$=80% |
| --- | --- | --- | --- | --- | --- | --- | --- | --- | --- |
| InAs (ML) | 9.68 | 8.82 | 7.90 | 6.90 | 5.90 | 4.90 | 3.90 | 2.90 | 1.90 |
| InSb (ML) | 0.32 | 0.19 | 0.10 | 0.10 | 0.10 | 0.10 | 0.10 | 0.10 | 0.10 |
| AlSb (ML) | - | 1.00 | 0.95 | 1.37 | 1.77 | 2.20 | 2.63 | 3.09 | 3.57 |
| AlAs (ML) | - | - | 0.10 | 0.26 | 0.46 | 0.60 | 0.74 | 0.83 | 0.87 |
| AlSb (ML) | - | - | 0.95 | 1.37 | 1.77 | 2.20 | 2.63 | 3.09 | 3.57 |

At first, supercells for Al$_x$In$_{1-x}$As$_y$Sb$_{1-y}$ digital alloy are taken in the form shown in Fig. 2, where $x$= 40%, 60% and 80% are shown as examples.

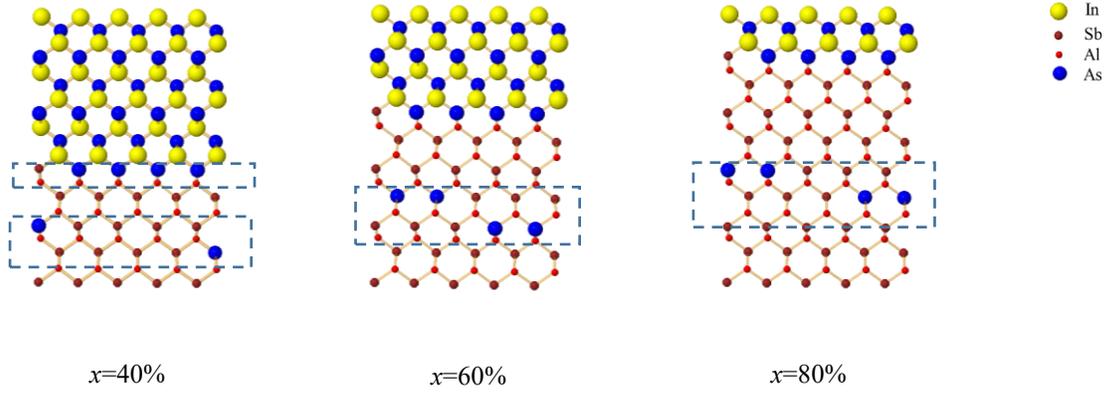

FIG. 2. Supercell for $Al_xIn_{1-x}As_ySb_{1-y}$ digital alloy with decimal number of monolayers in each period.

The supercell consists of 100 atoms. Fractional monolayers overlap with adjacent layers as illustrated in Fig. 2 with the dashed rectangular boxes.

Strain is also included in this model. The lattice constant, $a$, Poisson ratio, $D_i$, and shear moduli, $G_i$, are provided in Table II. Lateral and vertical lattice constants ($a_\parallel$ and $a_{i\perp}$) can be calculated by the following equations[8]

$$a_\parallel = \frac{a_1 G_1 h_1 + a_2 G_2 h_2}{G_1 h_1 + G_2 h_2} \quad (1)$$

$$a_{i\perp} = a_i [1 - D_i \varepsilon_i] \quad (2)$$

$$\varepsilon_i = a_\parallel / a_i - 1 \quad (3)$$

where $i$ denotes the adjacent material and $h_i$ is the layer thickness. From table I, it can be seen that in the $Al_xIn_{1-x}As_ySb_{1-y}$ digital alloy, each period is primarily composed of InAs and AlSb. Therefore, for simplification, the influence of InSb and AlAs on the lattice parameters is ignored.

TABLE II. Lattice constant $a$ (in Å), shear moduli $G$ and $D_i$ for InAs, InSb, AlAs and AlSb used in this work[13].

|      | $a$  | $D^{001}$ | $G^{001}$ | $D^{110}$ | $G^{110}$ | $D^{111}$ | $G^{111}$ |
|------|------|-----------|-----------|-----------|-----------|-----------|-----------|
| InAs | 6.08 | 1.088     | 1.587     | 0.674     | 2.306     | 0.570     | 2.487     |
| InSb | 6.48 | 1.080     | 1.261     | 0.698     | 1.785     | 0.600     | 1.920     |
| AlAs | 5.65 | 0.854     | 2.656     | 0.616     | 3.207     | 0.550     | 3.361     |
| AlSb | 6.08 | 0.990     | 1.763     | 0.641     | 2.372     | 0.550     | 2.530     |

E-k relationships along the [001] direction for $x$ varying from 0% to 80% have been calculated and $x$=0%, 40% and 80% are shown in Fig. 3 as examples. It has been found that the valence band maximum is relatively independent of $x$.

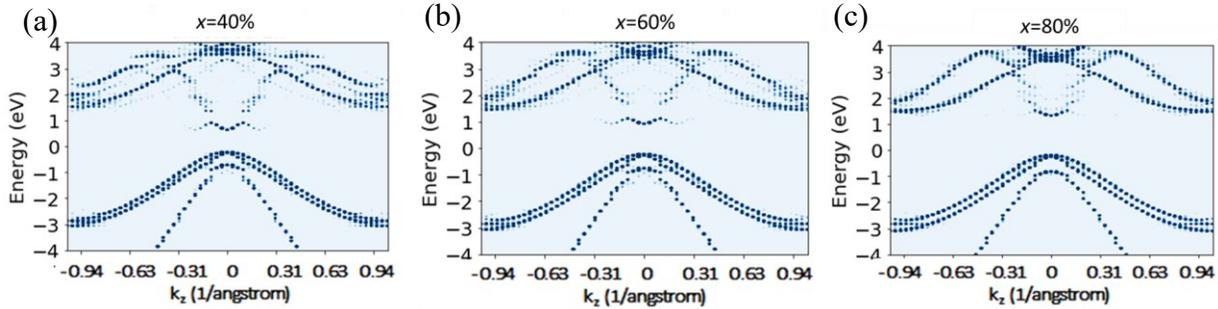

Fig. 3 E-k relationships along [001] direction for $Al_xIn_{1-x}As_ySb_{1-y}$ samples with (a) $x$=40%, (b) $x$=60% and (c) $x$=80%

The valence band maximum and conduction band minimum values are plotted as dashed and solid curves, respectively, in Fig. 4. From 0% to 80%, the increase in the valence band maximum is < 0.1 eV.

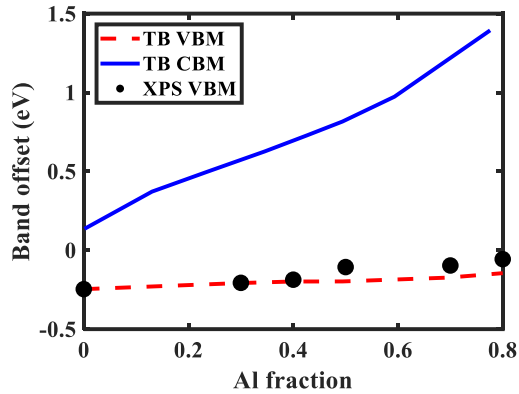

Fig. 4 Tight binding calculating results for valence band maximum (VBM) and conduction band minimum (CBM) and XPS measurement results for VBM of $Al_xIn_{1-x}As_ySb_{1-y}$ digital alloy.

In order to determine the effect of the lateral randomness caused by fractional monolayers in the supercell on band structure we considered the configuration with no overlap between adjacent layers, i.e., whole numbers of monolayers, as shown in Table III. The bandstructure of $Al_xIn_{1-x}As_ySb_{1-y}$ with $x$=60% was re-calculated. The super cell and E-k relationship are shown in Fig. 5. Comparing Fig. 5(b) with Fig. 3(b), the band structure doesn't change a lot. The valence band maximum shifts upward ~ 0.06 eV and the conduction band minimum shifts downward ~ 0.08 eV. We conclude that the lateral randomness caused by adjacent layer merging is relatively insignificant.

TABLE III. $Al_xIn_{1-x}As_ySb_{1-y}$ digital alloy layer structures in a unit cell for x varying from 0 to 80% with whole number of monolayers adding up to 10ML.

|          | x=0% | x=10% | x=20% | x=30% | x=40% | x=50% | x=60% | x=70% | x=80% |
|----------|------|-------|-------|-------|-------|-------|-------|-------|-------|
| InAs (ML)| 10   | 9     | 8     | 7     | 6     | 5     | 4     | 3     | 2     |
| InSb (ML)| -    | -     | -     | -     | -     | -     | -     | -     | -     |
| AlSb (ML)| -    | 1     | 1     | 1     | 1     | 2     | 2     | 3     | 4     |
| AlAs (ML)| -    | -     | -     | 1     | 1     | 1     | 1     | 1     | 1     |
| AlSb (ML)| -    | -     | 1     | 1     | 2     | 2     | 3     | 3     | 3     |

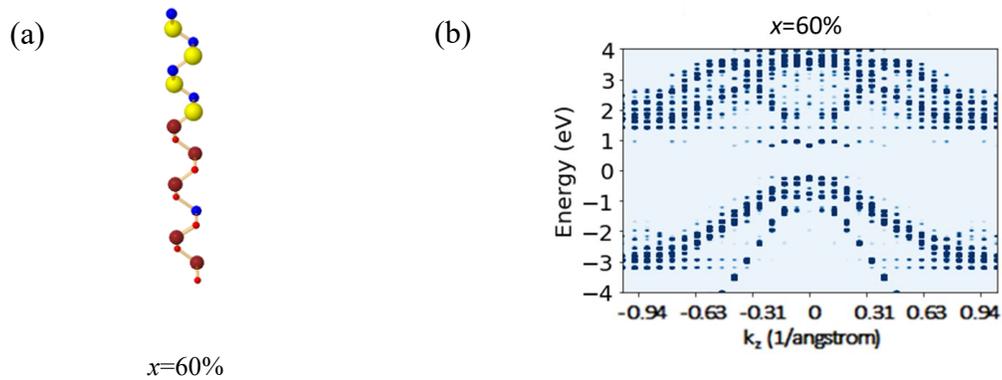

(a) x=60%

(b) x=60%

Fig. 5 Ultimate condition of lateral randomness (a) single column simplification of super cell with x=60% and (b) E-k relationship.

In order to confirm the $Al_xIn_{1-x}As_ySb_{1-y}$ digital alloy band offset with different *x*, X-ray photoelectron spectroscopy (XPS) measurements were carried out using a PHI VersaProbe III system with a monochromatic Al k-alpha X-ray source (1486.7 eV). The valence band maximum values were fixed at the intersection point of the linear extrapolation of the leading edge and the valence band spectra base line (a 0.05 eV spectrum resolution was used). In this work, Sb *4d* cathode-luminescence was selected to obtain the valence band offset, $\Delta E_V$, between $Al_{x1}In_{1-x1}As_{y1}Sb_{1-y1}$ and $Al_{x2}In_{1-x2}As_{y2}Sb_{1-y2}$ by using the Kraut model[14-16], which is given by the following equation, which calculates energy shifts upon altering compositions from $(Al_{x1}, As_{y1})$ to $(Al_{x2}, As_{y2})$ first in the Sb *4d* state and then in the VBM.

$$\Delta E_V = \left(E_{Sb_{4d}}^{Al_{x1}In_{1-x1}As_{y1}Sb_{1-y1}} - E_{VBM}^{Al_{x1}In_{1-x1}As_{y1}Sb_{1-y1}}\right)$$

$$- \left(E_{Sb_{4d}}^{Al_{x2}In_{1-x2}As_{y2}Sb_{1-y2}} - E_{VBM}^{Al_{x2}In_{1-x2}As_{y2}Sb_{1-y2}}\right) \quad (4)$$

$$- \left(E_{Sb_{4d}}^{Al_{x1}In_{1-x1}As_{y1}Sb_{1-y1}} - E_{Sb_{4d}}^{Al_{x2}In_{1-x2}As_{y2}Sb_{1-y2}}\right)$$

where $E_{Sb_{4d}}^{Al_xIn_{1-x}As_ySb_{1-y}}$ denotes the $Sb_{4d}$ peak in $Al_xIn_{1-x}As_ySb_{1-y}$, which can vary from sample to sample owing to differences in the surface potential, as shown in Fig. 6(a). Figure 6(b) shows the intersectional points for x varying from 0% to 80%. It is clear that the valence band maximum occurs at nearly the same energy for all compositions.

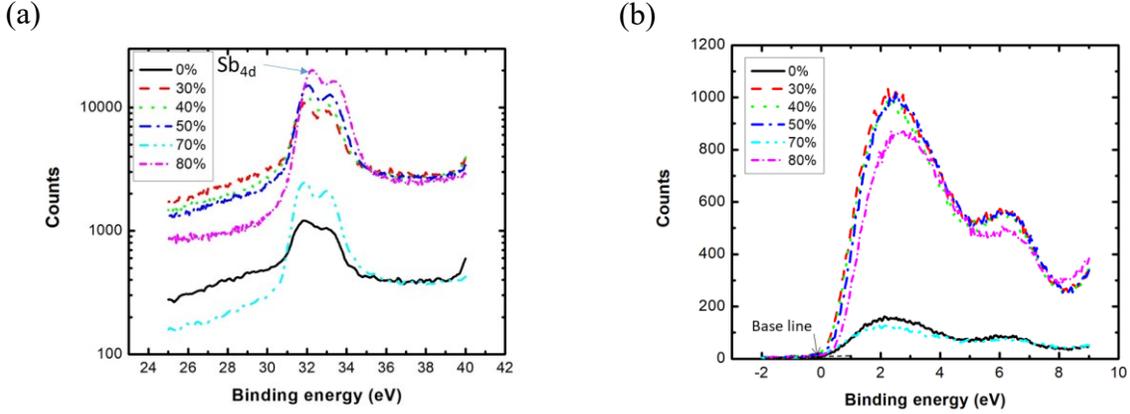

Fig. 6 XPS data of (a) Sb$_{4d}$ peaks and (b) VBM for Al$_x$In$_{1-x}$As$_y$Sb$_{1-y}$ with x varying from 0% to 80%.

The valance band offset for different $Al_xIn_{1-x}As_ySb_{1-y}$ compositions have been calculated and are plotted as in Fig. 4 as solid circles, •. Consistent with the band structure calculations, the valence band discontinuity is relatively independent of composition. It follows that the bandgap discontinuity is primarily in the conduction band.

The big variation in conduction band offset and small variation in valence band offset between the two binary constituent alloys (mainly between AlSb and InAs[17]) is consistent with observed minigaps inside the conduction band and their absence in the valence band (Fig. 3 and Fig. 5) and also the band offset variation between AlInAsSb with different Al compositions. A detailed transport model will be necessary hereafter to connect these band minigaps with the charge transmission, which depends on various details such as scattering potential, transport and tunneling effective masses and phonon energies.

In conclusion, the valence band offset between $Al_xIn_{1-x}As_ySb_{1-y}$ materials latticed matched to GaSb with x varying from 0 to 80% has been found to be nearly 0 by using tight binding calculations and XPS measurements. The change in bandgap energy with Al fraction, therefore, is primarily due to the conduction band offset.

This work has been supported by the Army Research Office (W911NF-17-1-0065) and DARPA (GG11972.153060).